\documentclass[codesnippetm, nojss]{jss}
\usepackage{thumbpdf,lmodern} 
\graphicspath{{Figures/}}

\newcommand{\R}{\proglang{R}~}
\newcommand{\C}{\proglang{C}~}
\newcommand{\Cpp}{\proglang{C++}~}
\newcommand{\RcppThread}{\pkg{RcppThread}~}

\author{Thomas Nagler\\Leiden University}
\title{\proglang{R}-Friendly Multi-Threading in \proglang{C++}}

\Plainauthor{Thomas Nagler} 
\Plaintitle{R-Friendly Multi-Threading in C++} 
\Shorttitle{\proglang{R}-Friendly Multi-Threading in \proglang{C++}} 

\Abstract{ Calling multi-threaded \Cpp code from \R has its
  perils. Since the \R interpreter is single-threaded, one must not
  check for user interruptions or print to the \R console from
  multiple threads. One can, however, synchronize with \R from the
  main thread. The \R package \RcppThread (current version 1.0.0)
  contains a header only \Cpp library for thread safe communication
  with \R that exploits this fact. It includes \Cpp classes for
  threads, a thread pool, and parallel loops that routinely
  synchronize with \proglang{R}. This article explains the package's
  functionality and gives examples of its usage. The synchronization
  mechanism may also apply to other threading frameworks.  Benchmarks
  suggest that, although synchronization causes overhead, the parallel
  abstractions of \RcppThread are competitive with other popular
  libraries in typical scenarios encountered in statistical computing. 
  A version of this vignette has been formally published as \citet{nagler21}.
}

\Plainkeywords{R, C++, parallel, thread, concurrency}

\Address{
  Thomas Nagler\\
  Leiden University \\
  Mathematical Institute \\
  Niels Bohrweg 1, 2333 CA Leiden, The Netherlands \\
  E-mail: \email{mail@tnagler.com}\\
  URL: \url{http://tnagler.com}
}

\begin{document}


\section{Introduction}

\subsection{From single to multi-cores machines}

For a long time, computers had only a single CPU and computer programs
were a set of instructions that the CPU executed in sequential order.
Accordingly, most programming languages that are still popular today
(including \proglang{R}, \citealt{R}, and \proglang{C++}) were
designed with a single-processor model in mind.  Computing power was
growing at exponential rates for decades and there was no reason to
change anything about that.

A paradigm shift came shortly after the turn of the millennium.
\citet{sutter2005} warned that the ``free lunch will soon be over'':
although the number of transistors on CPUs is continuing to grow
exponentially, their clock speed is approaching physical limits.  To
keep increasing the computing power, manufacturers made a move towards
multi-core machines.  Today, virtually all desktop PCs, laptops, and
even smart phones have multiple cores to deal with increasingly
demanding software applications.

As time progresses, statistical tools and methods are becoming
increasingly complex and demanding for the computer.  For that reason,
many \proglang{R} packages implement performance-critical tasks in a
lower level language like \proglang{C} and \proglang{Fortran}.
Interfacing with \proglang{C++} has become especially popular in
recent years thanks to the excellent \pkg{Rcpp} package \citep{Rcpp1,
  Rcpp2}, which is used by almost 1\,500 ($\approx 10\%$ of total)
packages on the Comprehensive \proglang{R} Archive Network (CRAN).

\subsection{Threads as programming abstraction for multi-core hardware}

To get the most out of a modern computer, it is vital to utilize not
only one, but many cores concurrently.  This can be achieved by
allowing multiple threads to be executed at the same time.  A thread
encapsulates a sequence of instructions that can be managed by a task
scheduler, typically the operating system.  A simple program has only
a single thread of execution, that encapsulates all instructions in
the program, the main thread.  However, a thread can also spawn new
threads that may run concurrently until all work is done and the
threads are joined. If a program contains more than one thread, we
speak of \emph{concurrency}.

In general, a single CPU can execute a program that has concurrent threads. The CPU may jump back and forth between the threads until all work is done (this is called context switching). However, the most gain in performance usually comes from \emph{parallelism}, i.e., when multiple CPUs work on multiple threads at the same time.

There are several frameworks that provide abstractions for running
multi-threaded code in \proglang{C++}.  The veteran among them is
\pkg{OpenMP} \citep{dagum1998openmp}, which provides preprocessor
directives to mark code sections that run concurrently.  More modern
frameworks include \pkg{Intel TBB} \citep{pheatt2008intel},
\pkg{Boost.Thread} \citep{BoostLibrary}, and \pkg{Tinythread++}
\citep{TinyThread}.  Since the advent of \proglang{C++11}, the
standard library provides low-level tools for managing concurrent and
parallel code. This includes the class `\code{std::thread}' that wraps a
handle to operating system threads, and tools for synchronization
between threads, like locks and atomic variables.

\subsection[Calling multi-threaded code from R]{Calling multi-threaded code from \proglang{R}}

Calling multi-threaded \proglang{C++} code from \proglang{R} can be
problematic because the \proglang{R} interpreter is single-threaded.
To quote from the `Writing \proglang{R} Extensions' manual
\citep[Section 1.2.1.1]{Rexts}: ``Calling any of the \R API from
threaded code is `for experts only'{}''.  Using \proglang{R}'s API from
concurrent threads may crash the \proglang{R} session or cause other
unexpected behavior.  In particular, communication between \Cpp code
and \R is problematic.  We can neither check for user interruptions
during long computations nor should we print messages to the
\proglang{R} console from any other than the main thread. It is
possible to resolve this, but not without effort.

\subsection[RcppThread and related packages]{\pkg{RcppThread} and related packages}

The \proglang{R} package \pkg{RcppThread} \citep{RcppThread} aims to
relieve package developers of that burden.  It contains \proglang{C++}
headers that provide:
\begin{itemize}
  \item thread safe versions of \code{Rcpp::Rcout} and \code{Rcpp::checkUserInterrupt()},
  \item parallel abstractions: thread, thread pool, and parallel \code{for} loops.
\end{itemize}
A word of caution: While \pkg{RcppThread} makes it possible to safely print from and interrupt multi-threaded \proglang{C++} code, all other parts of the \proglang{R} API remain unsafe and should be avoided.

\pkg{RcppThread}'s implementation only relies on built-in
\proglang{C++11} functionality for managing and synchronizing
threads. Hence, it only requires \proglang{C++11} compatible compiler
and is otherwise available for all operating systems. The package is
open source released with the MIT License and publicly available from
CRAN at \url{https://CRAN.R-project.org/package=RcppThread}.

Besides the numerous packages for parallelism in \proglang{R} (see
\citealt{HighPerformanceComputing-view}), there are two packages that
inspired \RcppThread and provide similar
functionality. \pkg{RcppProgress} \citep{RcppProgress} allows to
safely check for user interruptions when code is parallelized with
\pkg{OpenMP} (but only then). Further, \pkg{RcppParallel}
\citep{RcppParallel} is an interface to many high-level parallel
abstractions provided by \pkg{Intel TBB}, but does not allow for
thread safe communication with \proglang{R}.


\section[Thread safe communication with R]{Thread safe communication with \proglang{R}} \label{sec:communication}

It is not safe to call \proglang{R}'s \C API from multiple threads.
It is safe, however, to call it from the main thread.  That is the
idea behind \pkg{RcppThread}'s \code{checkUserInterrupt()} and
\code{Rcout}.  They behave almost like their \pkg{Rcpp} versions, but
only communicate with \R when called from the main thread.

\subsection{Interrupting computations}

\proglang{R} handles interruptions by internal signals that may
immediately terminate a computation.  Some IDEs (integrated
development environments), most notably RStudio \citep{RStudio}, wrap
around this behavior by setting a flag on the \proglang{R} session.
Whenever the \proglang{R} session encounters this flag, it sends a
termination signal and resets the flag. The \proglang{R} interpreter
checks for such flags often, such that pure \proglang{R} code can
terminated instantly.  However, \proglang{C++} routines do not benefit
automatically from this mechanism; developers must explicitly request
a check for user interruptions.  The \pkg{Rcpp} function
\code{checkUserInterrupt()} is a convenient way to request this check,
but it must not be called from child threads.

It is fairly easy to make \code{checkUserInterrupt()} thread safe. 
We first check whether the function is called from the main thread, and only then we ask \proglang{R} whether there was a user interruption. 

Consider the following example with `\code{std::thread}':
\begin{Code}
#include <RcppThread.h>
// [[Rcpp::export]]
void check()
{
  auto job = [] { RcppThread::checkUserInterrupt(); };
  std::thread t(job);
  t.join();
}
\end{Code}
The first line includes the \pkg{RcppThread} header, which
automatically includes the standard library headers required for
`\code{std::thread}' and `\code{std::chrono}'. The second line triggers
\pkg{Rcpp} to export the function to \proglang{R}. We define a
function \code{check()}. In the function body, we declare a function
\code{job()} that only checks for a user interruption. We then create
an `\code{std::thread}' with the new job and join it before the program
exits.

If we call the above function from \proglang{R}, the program completes
as expected.  But would we have used \code{Rcpp::checkUserInterrupt()}
instead, the program would terminate unexpectedly and crash the \R
session.

If \code{RcppThread::checkUserInterrupt()} is called from the main
thread and the user signaled an interruption, a
`\code{UserInterruptException}' will be thrown. This translates to an
error in \R with the message
\begin{CodeChunk}
\begin{CodeOutput}
C++ call interrupted by the user.
\end{CodeOutput}  
\end{CodeChunk}
A related function is \code{isInterrupted()} which does not throw an
exception, but returns a Boolean signaling the interruption
status. This can be useful, if some additional cleanup is necessary or
one wants to print diagnostics to the \R console.

However, when the functions are called from a child thread, they do
not actually check for an interruption.  This can be problematic if
they are only called from child threads. That does not happen with
\pkg{OpenMP} or \pkg{Intel TBB}, but with lower level frameworks like
`\code{std::thread}', \pkg{TinyThread++} or \pkg{Boost.Thread}.

Both functions accept a \code{bool} that allows to check conditionally
on the state of the program. For example, in a loop over \code{i},
\code{checkUserInterrupt(i \% 20 == 0)} will only check in every 20th
iteration. Checking for interruptions is quite fast (usually
microseconds), but there is a small overhead that can accumulate to
something significant in loops with many iterations. Checking
conditionally can mitigate this overhead.

There is a hidden detail worth mentioning. The two functions above are
not completely useless when called from a child thread. They check for
a global variable indicating whether the main thread has noticed an
interruption. Hence, as soon the main thread witnesses an
interruption, all child threads become aware.

In Section~\ref{sec:thread_impl}, we will discuss how to make sure that \code{isInterrupted()} is called from the main thread every now and then. 
For now, we are only able to write functions that are interruptable from the main thread, and safe to call from child threads.

\subsection[Printing to the R console]{Printing to the \proglang{R} console}

A similar issue arises when multiple threads try to print to the \R
console simultaneously. Consider the following example:
\begin{Code}
#include <thread>
#include <Rcpp.h>
// [[Rcpp::export]]
void greet()
{
  auto job = [] () {
    for (size_t i = 0; i < 100; ++i) 
    Rcpp::Rcout << "Hi!" << std::endl;
  };
  std::thread t1(job);
  std::thread t2(job);
  t1.join();
  t2.join();
}
\end{Code}
We create a \code{job} function that prints the message \code{"Hi!"}
to the \R console 100 times. We spawn two threads that execute the
job, and join them before the program exits. We expect the function to
print a stream of 200 messages saying \code{"Hi!"} in the \R
console. We can get lucky, but normally the two threads will try to
say \code{"Hi!"} at least once at the same time. Again, the \R session
would terminate unexpectedly.

Now consider the following variant:
\begin{Code}
#include <RcppThread.h>
// [[Rcpp::export]]
void greet()
{
  auto job = [] () {
    for (size_t i = 0; i < 100; ++i) 
    RcppThread::Rcout << "Hi!" << std::endl;
  };
  std::thread t1(job);
  std::thread t2(job);
  t1.join();
  t2.join();
  RcppThread::Rcout << "";
}
\end{Code}
This function will print 200 messages in the \R console as expected. 
But \code{RcppThread::Rcout} never prints to the console from child threads, so how does this work?

\code{RcppThread::Rcout} does not print to the \R console directly. It
stores the message in global buffer that is protected by a lock. Then
it checks whether it was called from the main thread. If this is not
the case, it does nothing further. If it was called from the main
thread, it releases all messages that are currently in the
buffer. Notice that we print an empty message from the main thread in
the last line of the program. This ensures that all messages are
released from the buffer before the program exits.


\section[An R-friendly thread class]{An \proglang{R}-friendly thread class} \label{sec:thread}

As of \proglang{C++11}, the standard template library provides the
class `\code{std::thread}' for executing code sections concurrently. The
implementation and syntax are very similar to \pkg{Boost.Thread} and
\pkg{TinyThread++}. \pkg{RcppThread}'s `\code{Thread}' class is an
\proglang{R}-friendly wrapper to `\code{std::thread}'.

Instances of class `\code{Thread}' behave almost like instances of
`\code{std::thread}'. There is one important difference: Whenever child
threads are running, the main thread periodically synchronizes with
\proglang{R}. In particular, it checks for user interruptions and
releases all messages passed to \code{RcppThread::Rcout}. When the
user interrupts a threaded computation, any thread will stop as soon
it encounters \code{checkUserInterrupt()}.

\subsection{Functionality}

Let us start with an example:
\begin{Code}
#include <RcppThread.h> 
// [[Rcpp::export]]
void pyjamaParty()
{
  using namespace RcppThread;
  auto job = [] (int id) {       
    std::this_thread::sleep_for(std::chrono::seconds(1));
    Rcout << id << " slept for one second" << std::endl;
    
    checkUserInterrupt();
    
    std::this_thread::sleep_for(std::chrono::seconds(1));
    Rcout << id << " slept for another second" << std::endl;
  };
  
  Thread t1(job, 1);
  Thread t2(job, 2);
  t1.join();
  t2.join();
}
\end{Code}
We create a function \code{job} that takes an integer \code{id} as argument
and does the following: Sleep for one second, send a message, check
for a user interruption, go back to sleep, and send another
message. We spawn two new `\code{Thread}'s with this job and join the
threads before the program exits. Notice that the argument of the
\code{job} function is passed to the `\code{Thread}' constructor. More
generally, if a job function takes arguments, they must be passed to
the constructor as a comma-separated list.

The example from the previous section used `\code{std::thread}' and was
not interruptible. The reason is that \code{checkUserInterrupt()} was
only called from child threads. This example is similar. However, the
`\code{Thread}' objects synchronize with \R and periodically check for
user interruptions. If we call the function from \R and interrupt the
computation, we get the following.
\begin{CodeChunk}
\begin{CodeInput}
> pyjamaParty()
\end{CodeInput}
\begin{CodeOutput}
1 slept for one second
2 slept for one second
Error in pyjamaParty() : C++ call interrupted by user
\end{CodeOutput}   
\end{CodeChunk}
The execution is interrupted after the two threads were done with the
first round of sleep. Further, although there was no \code{Rcout}
statement in the main thread, the messages were sent to the \R
console. The `\code{Thread}' instances took care of both checking for
interruptions and releasing messages to the \R console.

The two \code{.join()} statements are important in this
example. Threads should always be joined before they are
destructed. The \code{.join()} statements signal the main thread to
wait until the jobs have finished. But instead of just waiting, the
main thread starts synchronizing with \proglang{R}.

The class `\code{Thread}' also allows for all additional functionality
(like swapping or detaching) provided by `\code{std::thread}'. However,
detaching threads that communicate with \proglang{R} should generally
be avoided.

\subsection{Implementation} \label{sec:thread_impl}

The synchronization mechanism bears some interest because it can be
implemented similarly for threading frameworks other than
`\code{std::thread}'. The foundation is a concept called future.  A
future allows to start a side-task and continue with the program,
until -- at some later point in time -- we explicitly request the
result.

Let us first have a look at a simplified version of the `\code{Thread}'
class constructor.
\begin{Code}
template<class Function, class... Arguments>
Thread(Function&& f, Arguments&&... args)
{
  auto f0 = [=] { f(args...); };
  auto task = std::packaged_task<void()>(f0);
  future_ = task.get_future();
  thread_ = std::thread(std::move(task));
}
\end{Code}
The constructor is a variadic template that takes a function and an
arbitrary number of additional arguments. The function \code{f} should
be a callable object and the additional arguments such that
\code{f(args...)} is a valid call. The constructor creates a new
function \code{f0} that evaluates \code{f}, passing it all additional
arguments (if there are any). The new function \code{f0} is wrapped in
an `\code{std::packaged_task}' that allows to access the result by a
future.  The future is stored in a class member \code{future_} and the
task is run in an `\code{std::thread}'.

The synchronization mechanism is in \code{join()}:
\begin{Code}
void join()
{
  auto timeout = std::chrono::milliseconds(250);
  while (future_.wait_for(timeout) != std::future_status::ready) {
    Rcout << "";
    if (isInterrupted())
    break;
  }
  thread_.join();
  Rcout << "";
  checkUserInterrupt();
}
\end{Code}
The function runs a \code{while} loop that relies on the future. The
condition of the loop lets the main thread sleep until one of two
events occur. One event is that a timeout of 250ms has been
reached. After waking up, the thread releases all messages to the \R
console and checks for an interruption. If there was an interruption,
the call to \code{isInterrupted()} will set the global flag for
interruption (so child threads become aware) and exit the
loop. Otherwise, the \code{while} loop continues and the main thread
again waits for one of the two events. The second event is that the
result of \code{f(args...)} is available. The \code{while} loop exits
and the internal `\code{std::thread}' object is joined. We again release
all messages in the buffer and call \code{checkUserInterrupt()}. The
latter ensures that an exception is thrown if there was a user
interruption.

The choice of 250ms for the timeout is somewhat arbitrary. It is short
enough to avoid long waiting times for an interrupting user. At the
same time, it is long enough such that any overhead from
synchronization becomes negligible.


\section{Parallel abstractions} \label{sec:abstractions} When there
are more than a few jobs to run, plain threads can be tedious.  Every
job requires spawning and joining a thread. This has a small, but
non-negligible overhead. Even worse: If there are more threads than
cores, the program may actually slow down. The \RcppThread package
provides two common abstractions to make life easier. Both synchronize
with \R using a similar mechanism as `\code{Thread}'.

\subsection{An interruptible thread pool}

A thread pool consists of a task queue and a fixed number of worker
threads.  Whenever the task queue contains jobs, an idle worker
fetches one task and does the work. Besides ease of use, the thread
pool pattern has two benefits. Tasks are assigned to workers
dynamically, so all workers are kept busy until there are no tasks
left. This is especially useful when some tasks take more time than
others. The second benefit is that one can easily limit the number of
threads running concurrently.

\subsubsection{Basic usage}

The class `\code{ThreadPool}' implements the thread pool pattern in a
way that plays nicely with \code{checkUserInterrupt()} and
\code{Rcout}. Its usage is fairly simple.
\begin{Code}
ThreadPool pool(3);
std::vector<int> x(100);
auto task = [&x] (unsigned int i) { x[i] = i; };
for (unsigned int i = 0; i < x.size(); ++i)
  pool.push(task, i);  
pool.join();
\end{Code}
The first line creates a `\code{ThreadPool}' object with three worker
threads. If the argument is omitted, the pool will use as many worker
threads as there are cores on the machine.  A thread pool initialized
with zero workers will do all work in the main thread. This makes it
easy to let the user decide whether computations run in parallel.

The second line instantiates a vector \code{x} to be filled in
parallel. The task function takes an index argument \code{i} and
assigns it to the \code{i}th element of \code{x}. The thread pool
knows about \code{x} because the lambda function captures its address
(expressed through \code{[\&x]}). We push 100 tasks to the thread
pool, each along with a different index. Then the thread pool is
joined. Again, the \code{join()} statement is important. First and
foremost, it causes the main thread to halt until all tasks are
done. But similar to \code{Thread::join()}, the thread pool starts to
periodically synchronize with \proglang{R}. Only after all work is
done, the worker threads in the pool are joined.

In the example, multiple threads write to the same object
concurrently. Generally, this is dangerous. In our example, however,
we know that the threads are accessing different memory locations,
because each task comes with a unique index. Code that writes to an
address that is accessed from another thread concurrently needs extra
synchronization (for example using a mutex). Read operations are
generally thread safe as long as nobody is writing at the same time.

The thread pool is interruptible without any explicit call to
\code{checkUserInterrupt()}. Before a worker executes a task, it
always checks for a user interruption.

\subsubsection{Tasks returning a value}

In some use cases, it is more convenient to let the tasks assigned to
the pool return a value. The function \code{pushReturn()} returns a
`\code{std::future}' to the result of the computations. After all jobs
are pushed, we can call \code{get()} on the future object to retrieve
the results:
\begin{Code}
ThreadPool pool;
auto task = [] (int i) { 
  double result;
  // some work
  return result;
};
std::vector<std::future<double>> futures(10);
std::vector<double> results(10);
for (unsigned int i = 0; i < 10; ++i)
  futures[i] = pool.pushReturn(task, i);
for (unsigned int i = 0; i < 10; ++i)
  results[i] = futures[i].get();
pool.join();
\end{Code}
\subsubsection{Using the same thread pool multiple times}

It is also possible to wait for a set of tasks and re-use the thread
pool by calling \code{wait()} instead of \code{join()}. The call to
\code{wait()} synchronizes with \R while waiting for the current jobs
to finish, but does not join the worker threads. When all tasks are
done, we can start pushing new jobs to the pool.

\subsection{Parallel for loops}

\subsubsection{Index-based parallel for loops}

The previous example used the thread pool to implement a very common
parallel pattern: a parallel for loop. The single-threaded version is
much simpler.
\begin{Code}
std::vector<int> x(100);
for (unsigned int i = 0; i < x.size(); ++i)
  x[i] = i;
\end{Code}
Since this pattern is so common, \RcppThread provides a wrapper
\code{parallelFor} that encapsulates the boiler plate from the thread
pool example. A parallel version of the above example can be expressed
similarly.
\begin{Code}
std::vector<int> x(100);
parallelFor(0, x.size(), [&x] (unsigned int i) {
  x[i] = i;
});
\end{Code}
There are differences between the single- and multi-threaded
version. The single-threaded version instantiates the loop counter in
the \code{for} declaration. The multi-threaded version, passes the
start and end indices and a lambda function that captures \code{\&x}
and takes the loop counter as an argument. The parallel version is a
bit less flexible regarding the loop's break condition and increment
operation. Additionally, the multi-threaded version may need
additional synchronization if the same memory address is accessed in
multiple iterations.

\subsubsection{Parallel for-each loops}

Another common pattern is the for-each loop. It loops over all elements in a container and applies the same operation to each element. A single-threaded example of such a loop is the following.
\begin{Code}
std::vector<int> x(100, 1);
for (auto& xx : x)
  xx *= 2;
\end{Code}
The \code{auto} loop runs over all elements of \code{x} and multiplies them by two.
The parallel version is similar:
\begin{Code}
std::vector<int> x(100, 1);
parallelForEach(x, [] (int& xx) {
  xx *= 2;
});
\end{Code}
Both \code{parallelFor} and \code{parallelForEach} use a
`\code{ThreadPool}' object under the hood and, hence, periodically
synchronize with \proglang{R}.

\subsubsection{Fine tuning the scheduling system}

The two functions \code{parallelFor} and \code{parallelForEach} essentially create a thread pool, push the tasks, join the pool, and exit. By default, there are as many worker threads in the pool as there are cores on the machine. The number of workers can be specified manually, however. The following example runs the loop from the previous example with only two workers (indicated by the \code{2} in the last line).
\begin{Code}
std::vector<int> x(100, 1);
parallelForEach(x, [] (int& xx) {
  xx *= 2;
}, 2);
\end{Code}
The syntax for \code{parallelFor} is similar.

There is more: The \code{parallelFor} and \code{parallelForEach}
functions bundle a set of tasks into batches. This can speed up code
significantly when the loop consists of many short-running iterations;
see, e.g., Figure~\ref{fig:benchEmpty} in
Section~\ref{sec:benchmarks}. Synchronization between worker threads in
the pool causes overhead that is reduced by packaging tasks into
batches. At the same time, we benefit from dynamic scheduling whenever
there are more batches than tasks. \RcppThread relies on heuristics to
determine an appropriate batch size. Sometimes, the performance of
loops can be improved by a more careful control over the batch
size. The two functions take a fourth argument that allows to set the
number of batches. The following code runs the loop with two workers
in 20 batches.
\begin{Code}
std::vector<int> x(100, 1);
parallelForEach(x, [] (int& xx) {
  xx *= 2;
}, 2, 20);
\end{Code}
\subsubsection{Calling for loops from a thread pool}

The functions create and join a thread pool every time they are called.  To reduce overhead, the functions can also be called as methods of a thread pool instance.
\begin{Code}
ThreadPool pool;
pool.parallelForEach(x, [] (int& xx) {
  xx *= 2;
});
pool.wait();
pool.parallelFor(0, x.size(), [&x] (int i) {
  x[i] *= 2;
});
pool.join();
\end{Code}
\subsubsection{Nested for loops}

Nested loops appear naturally when operating on multi-dimensional arrays. One can also nest the parallel for loops mentioned above. Although not necessary, it is more efficient to use the same thread pool for both loops. 
\begin{Code}
ThreadPool pool;
std::vector<std::vector<int>> x(100);
for (auto &xx : x)
  xx = std::vector<int>(100, 1.0);
pool.parallelFor(0, x.size(), [&] (int i) {
  pool.parallelFor(0, x[i].size(), [&, i] (int j) {
    x[i][j] *= 2;
  });
});
pool.join();
\end{Code}
The syntax for nested \code{parallelForEach} is similar. 

A few warnings: It is usually more efficient to run only the outer
loop in parallel. To minimize overhead, one should parallelize at the
highest level possible. Furthermore, if both inner and outer loops run
in parallel, we do not know the execution order of tasks. We must not
parallelize nested loops in which order matters. Captures of lambda
functions (or other callable objects replacing the loop body) require
extra care: Since the outer loop index \code{i} is temporary, it must
be copied.


\section[Using the RcppThread package in other projects]{Using the \proglang{RcppThread} package in other projects} \label{sec:projects}

The \RcppThread package contains a header-only \proglang{C++} library
that only requires a \proglang{C++11} compatible compiler. This is
only a mild restriction, because the standard has long been
implemented by most common compilers. The package has no dependencies,
although using \pkg{Rcpp} is strongly encouraged (the \pkg{Rcpp} layer
automatically converts `\code{UserInterruptException}' into an
\proglang{R} error messages). To use the package in other \R projects,
users only need to include the \RcppThread headers and enable
\proglang{C++11} functionality. In the following, we briefly explain
how this can be achieved.

\subsection[Using RcppThread in inline C++ snippets]{Using \RcppThread in inline \Cpp snippets}

The \code{cppFunction()} and \code{sourceCpp()} functions of the
\pkg{Rcpp} package provide a convenient way to quickly implement a
\proglang{C++} function inline from \proglang{R}. The following is a
minimal example of \RcppThread used with \code{cppFunction()}:
\begin{CodeChunk}
\begin{CodeInput}
R> func <- Rcpp::cppFunction("void func() { /* actual code here */ }", 
+    depends = "RcppThread", plugins = "cpp11")
\end{CodeInput}    
\end{CodeChunk}
The first argument of \code{cppFunction()} is a \proglang{C++} snippet
defining a function. The \code{depends = "RcppThread"} argument takes
care that the relevant \pkg{RcppThread} headers are included;
\code{plugins = "cpp11"} tells the compiler to enable \proglang{C++11}
functionality.  After running this line in the \R console,
\code{func()} can be called as a regular \R function.

The same can be achieved using \code{sourceCpp()}:
\begin{CodeChunk}
\begin{CodeInput}
R> Rcpp::sourceCpp(code = '// [[Rcpp::plugins(cpp11)]]
+    // [[Rcpp::depends(RcppThread)]]    
+    #include "RcppThread.h"
+    // [[Rcpp::export]]
+    void func() { /* actual code here */}')
\end{CodeInput}    
\end{CodeChunk}
Inside the \code{code} block, the first line enables \proglang{C++11},
the second tells the compiler where to look for \RcppThread headers,
which are included in the third line. The fourth and fifth lines
define the function and request it to be exported to \proglang{R}. The
\code{sourceCpp()} version is more verbose, but offers additional
flexibility, since other functions or classes can be defined in the
same code block.

\subsection[Using RcppThread in another R package]{Using \RcppThread in another \R package}

Using \RcppThread in other \R packages is similarly easy:
\begin{enumerate}
\item Add \code{CXX_STD = CXX11} to the \code{src/Makevars(.win)} files of your package.
\item Add \code{RcppThread} to the \code{LinkingTo} field in the \code{DESCRIPTION}.
\item Include the headers with \code{\#include "RcppThread.h"}.
\end{enumerate}

\subsection[Using RcppThread to port existing C++ code]{Using \RcppThread to port existing \Cpp code}

For packages porting existing \Cpp libraries to \proglang{R},
\RcppThread provides two preprocessor macros for convenience. The
respective \code{\#define} statements need to be placed before the
\RcppThread headers.

\begin{itemize}
\item \code{\#define RCPPTHREAD_OVERRIDE_COUT 1}: replaces all
  instances of \code{std::cout} with \\ \code{RcppThread::Rcout}.
\item \code{\#define RCPPTHREAD_OVERRIDE_THREAD 1}: replaces all
  instances of `\code{std::thread}' with `\code{RcppThread::Thread}'.
\end{itemize}


\section{Benchmarks} \label{sec:benchmarks}

Parallel computation is primarily about speed, so it is a good idea to
measure. In a first step, we want to quantify the overhead by
\pkg{RcppThread}'s synchronization with \proglang{R}.  The second part
compares the performance of the `\code{ThreadPool}' and
\code{parallelFor} abstractions against implementations based on
\pkg{RcppParallel} and \pkg{OpenMP}.

Results of computing benchmarks depend strongly on the hardware,
especially for parallel programs.  The results in this section were
recorded on an i5-6600K CPU with four cores at 3.5 GHz, 6MB cache
size, and 16GB 3GHz DDR4 RAM.  The code for the benchmarks is
available in the supplementary material, so readers can test them on
their own machine. 

\subsection{Synchronization overhead}

\subsubsection{Create, join, and destruct threads}

\begin{figure}[t!]
  \centering
  \includegraphics[width = 0.65\textwidth]{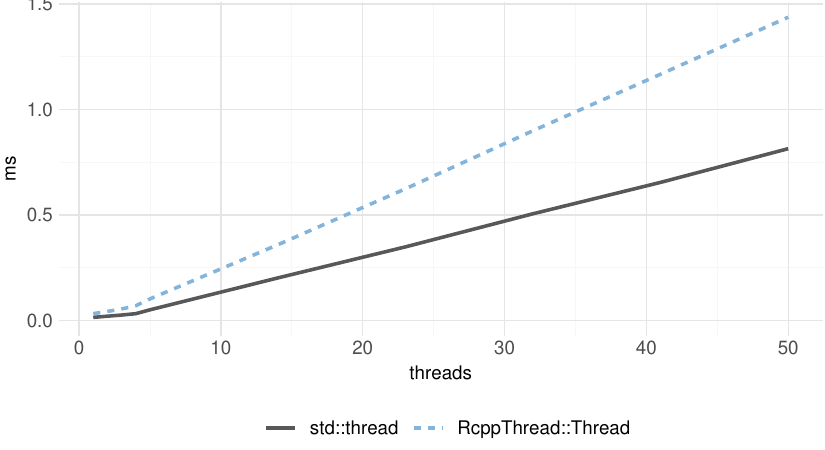}
  \caption{Time required for creating, joining, and destroying thread objects of class `\code{std::thread}' and `\code{RcppThread::Thread}'.}
  \label{fig:benchEmptyThread}
\end{figure}

As explained in Section~\ref{sec:thread_impl},
`\code{RcppThread::Thread}' encapsulates a `\code{std::thread}' object,
but exploits a `\code{std::future}' for additional synchronization. To
quantify the overhead, our first example simply creates a number of
thread objects, and then joins and destroys them.

The speed using `\code{RcppThread::Thread}' (dashed) is compared against
`\code{std::thread}' (solid) in Figure~\ref{fig:benchEmptyThread}. We
observe that `\code{RcppThread::Thread}' is roughly two times slower
than `\code{std::thread}'. Both lines show a kink at four threads. This
corresponds to the four physical cores on the benchmark machine. If
there are more threads than cores, we pay an additional fee for
`context switching' (jumping between threads).

The marginal cost of a single `\code{RcppThread::Thread}' is around
10$\mu$s when there are less than four threads, and roughly 30$\mu$s
otherwise. Although even 30$\mu$s sounds cheap, running more threads
than cores will also slow all other computations down. Thread pools or
parallel loops use a fixed number of threads and should be preferred
over naked threads when possible.

\subsubsection{Checking for user interruptions}

\begin{figure}[t!]
  \centering
  \includegraphics[width = 0.65\textwidth]{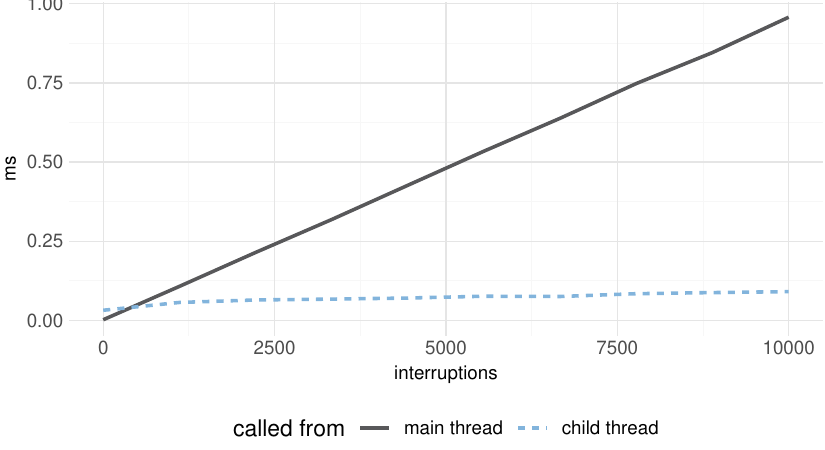}
  \caption{Time required to check for user interruptions from either the main or a child thread.}
  \label{fig:benchInterrupt}
\end{figure}

Figure~\ref{fig:benchInterrupt} shows the time it takes to call
\code{checkUserInterrupt()} either from the main or a child
thread. Checking for user interruptions is rather cheap: One check
costs around 100ns from the main thread, and 5ns from a child thread.
The latter is cheaper because there is no direct synchronization with
\proglang{R}. If called from a child thread,
\code{checkUserInterrupt()} only queries a global flag. Hence, we can
generously sprinkle parallelized code sections with such checks
without paying much in performance.

\subsection{Comparison to other parallel libraries}

\subsubsection{Empty jobs}

\begin{figure}[t!]
\centering
\includegraphics[width = 0.75\textwidth]{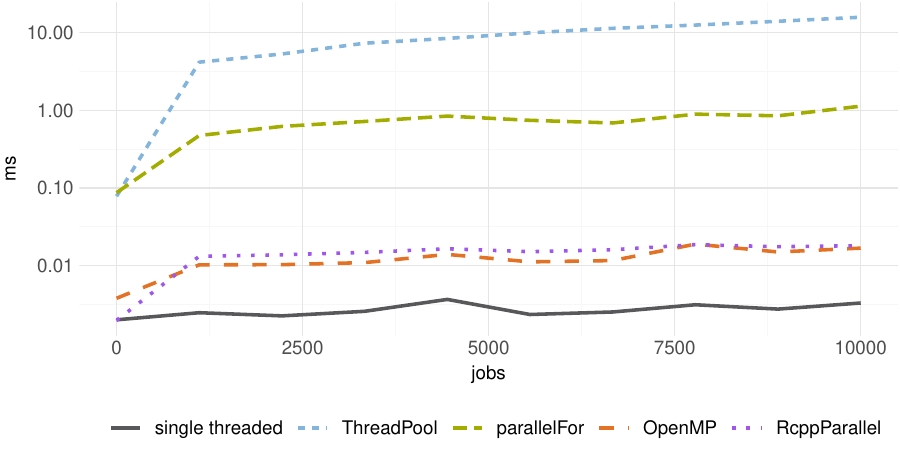}
\caption{Time required for submitting empty jobs to different parallelism frameworks.}
\label{fig:benchEmpty}
\end{figure}

To start, we measure the performance for running empty jobs. The solid
line in Figure~\ref{fig:benchEmpty} indicates the time required to run
a single threaded loop with \code{jobs} iterations doing nothing. The
other lines show the performance of various parallel abstractions:
\pkg{RcppThread}'s `\code{ThreadPool}' and \code{parallelFor}, and
parallel for loops based on \pkg{OpenMP} and \pkg{RcppParallel}.

The abstractions provided by \RcppThread are much slower than their
competitors (note the log scale on the $y$-axis). This has two
reasons. The \RcppThread functions are weighted with infrastructure
for synchronization with \proglang{R}. Further, the competitor
libraries are highly optimized for high-throughput scenarios by
avoiding memory locks as much as possible.

We also observe that the parallel for loops are much faster than the
thread pool. Since the thread pool accepts new jobs at any time, it
must handle any job as a separate instance. Parallel for loops know
upfront how much work there is and bundle jobs into a smaller number
of batches. This technique reduces the necessary amount of
synchronization between threads.

The single threaded version was the fastest, by far. Of course, we
cannot expect any gain from parallelism when there is nothing to
do. When jobs are that light, the overhead vastly outweighs the
benefits of concurrent computation.

\subsubsection{Computing kernel density estimates}

\begin{figure}[t!]
\centering
\includegraphics[width = \textwidth]{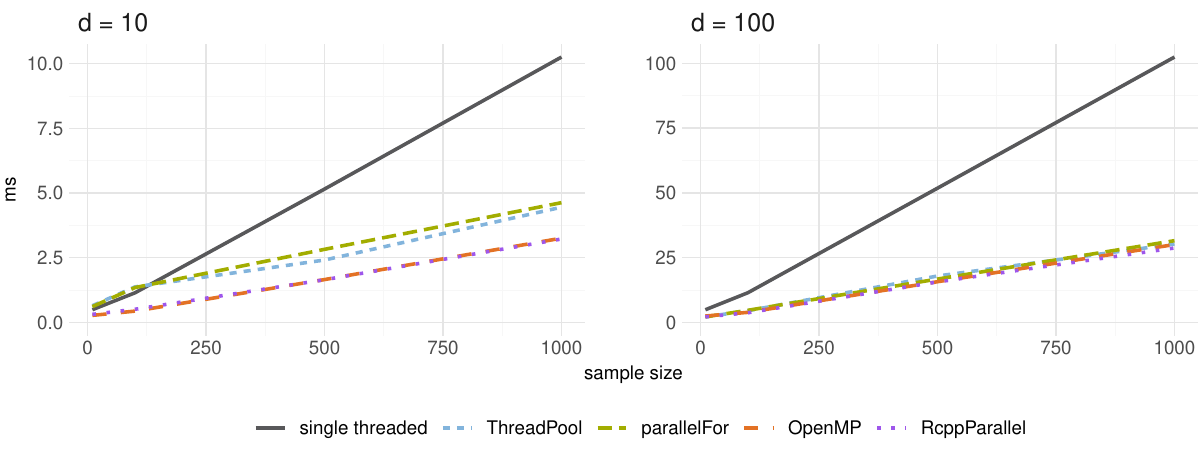}
\caption{Time required for computing the kernel density estimate in parallel for $d$ variables.}
\label{fig:benchKDE}
\end{figure}

Let us consider a scenario that is more realistic for statistical
applications. Suppose we observe data from several variables and want
to compute a kernel density estimate for each variable. This is a
common task in exploratory data analysis or nonparametric statistical
learning (e.g., the naive Bayes classifier) and is easy to
parallelize. For simplicity, the estimator is evaluated on 500 grid
points.

Figure~\ref{fig:benchKDE} shows the performance for $d = 10$ (left
panel), $d = 100$ variables (right panel), and increasing sample
size. For $d = 10$ and moderate sample size the two \RcppThread
functions are about 10\% slower than their competitors, but catch up
slowly for large samples. The shift is essentially the overhead we
measured in the previous benchmark. For $d = 100$, the overhead of
\RcppThread is negligible and all methods are on par. Generally, all
parallel methods are approximately 4$\times$ faster than the single
threaded version.

\subsubsection{Computing Kendall's correlation matrix}

\begin{figure}[t!]
  \centering
  \includegraphics[width = \textwidth]{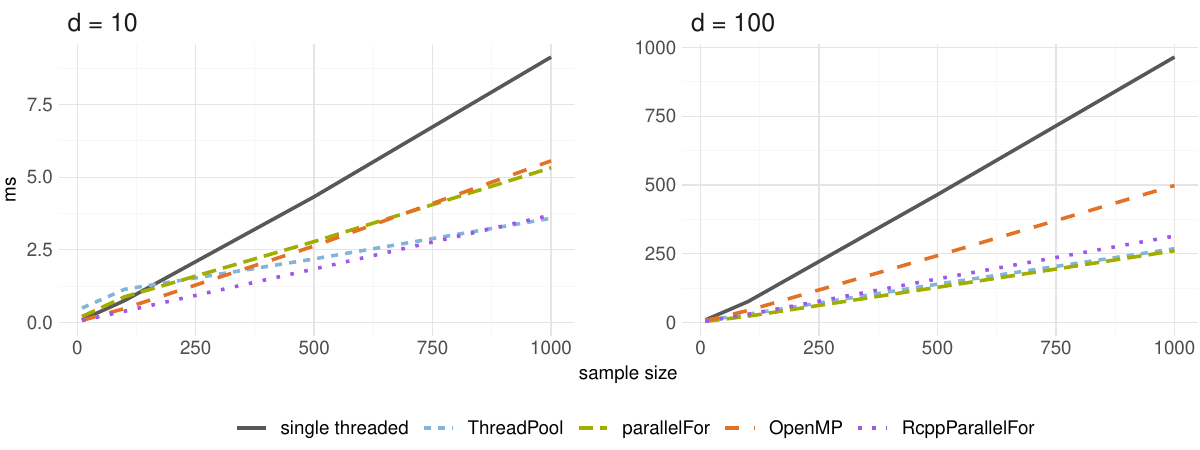}
  \caption{Time required for computing Kendall's correlation matrix in parallel for $d$ variables.}
  \label{fig:benchKendall}
\end{figure}

Now suppose we want to compute a matrix of pair-wise Kendall's
correlation coefficients. Kendall's $\tau$ is a popular rank-based
measure of association. In contrast to Pearson's correlation, it
measures monotonic (not just linear) dependence, but is
computationally more complex; \proglang{R}'s implementation in
\code{cor()} scales quadratically in the sample size
$n$. \citet{knight1966computer} proposed an efficient algorithm based
on a merge sort that scales $n \log n$ \citep[as implemented
by][]{wdm}. As a downside, the correlation matrix can no longer be
computed with matrix algebra; each coefficient $\tau_{i, j}$ must be
considered separately. There are ${d \choose 2}$ unique pairs of
variables $(i, j)$, $1 \le i < j \le d$. The coefficients are computed
in a nested loop over $i$ and $j$, where we only parallelize the outer
loop over $i$.

This problem is quite different from the kernel density estimation
benchmark. First, the problem scales quadratically in the dimension
$d$. And more importantly, the jobs are unbalanced: For each $i$,
there are only $d - i$ iterations in the inner loop. Hence, iterations
with small $i$ take longer than iterations with large $i$. The larger
the dimension $d$, the larger the imbalance.

Figure~\ref{fig:benchKendall} shows the benchmark results. For
$d = 10$, we observe that none of the parallel methods achieve a
4$\times$ speed-up. The reason is that the tasks are still rather
small. Even for $n = 1\,000$, each iteration of the inner loop takes
only a fraction of a millisecond. For such jobs, the parallelization
overhead becomes visible.  \code{parallelFor} is slowest among all
methods. For sample sizes smaller than 500, it is hardly faster than
the single threaded loop. Also \pkg{OpenMP} achieves less than a
2$\times$ improvement. Only `\code{ThreadPool}' and \pkg{RcppParallel}
achieve an approximate 3$\times$ speed up. Their scheduling appears to
better compensate the imbalance of the problem.

For $d = 100$, the picture is quite different. The \RcppThread
functions are faster than their competitors: roughly twice as fast as
\pkg{OpenMP} and $10\%$ faster than \pkg{RcppParallel}. Furthermore,
it gives an approximate 4$\times$ speed up, indicating an optimal use
of resources.

\subsection{Conclusions}

We conclude that the parallel abstractions provided by \RcppThread
cause notable overhead when concurrent tasks are small. For many
applications in statistical computing, however, this overhead becomes
negligible. In the future, the implementation \RcppThread may benefit
from additional optimizations. In particular, a lock free
implementation of the task queue may allow to reduce the overhead on
small tasks.  In any case, the main advantage is automatic and safe
synchronization with \proglang{R}, i.e., usability and not speed.

\bibliography{ref}

\end{document}